\documentclass[twocolumn,showpacs,amsmath,amssymb]{revtex4-1}
\usepackage{bm}
\usepackage[T1]{fontenc}
\usepackage{hyperref}
\input{epsf}

\usepackage{graphicx}
\usepackage{epstopdf}
\usepackage{array}
\usepackage{longtable}
\usepackage{rotating,booktabs}
\usepackage{booktabs,threeparttable}
\usepackage{bm}
\usepackage{float}
\usepackage{amsmath}
\usepackage{gensymb}
\usepackage{multirow}
\usepackage{epsfig}
\usepackage{color}

\begin{document}

\title{Dynamic polarizabilities and triple magic trapping conditions for $5s^2~^1S_0\rightarrow 5s5p~^3P_{0,2}$ transitions of Cd atoms}
\author{Ru-Kui Zhang$^{1,2}$}
\author{Jun Jiang$^{1}$}
\email {phyjiang@yeah.net}
\author{Chen-Zhong Dong$^{1}$}
\author{Yong-Bo Tang$^{2}$}
\email {tangyongbo@sztu.edu.cn}

\affiliation{$^{1}$Key Laboratory of Atomic and Molecular
	Physics and Functional Materials of Gansu Province,
	College of Physics and Electronic Engineering,
	Northwest Normal University, Lanzhou 730070, P. R. China}

\affiliation{$^{2}$Physics Teaching and Experiment Center, Shenzhen Technology University, Shenzhen 518118, P. R. China}

\date{\today}

\begin{abstract}
The dynamic electric dipole polarizabilities of the $5s^2~^1S_0$, $5s5p~^3P_{0}$, and $5s5p~^3P_2$ states for Cd atoms are calculated using the relativistic configuration interaction plus many-body perturbation theory method. The magic wavelengths for the $5s^2~^1S_0\rightarrow 5s5p~^3P_{0}$ and $5s^2~^1S_0\rightarrow 5s5p~^3P_2$ transitions within a range of 300-500 nm are identified.  The possibility of achieving triple magic trapping for the transitions $5s^2~^1S_0\rightarrow5s5p~^3P_{0}$ and $5s^2~^1S_0\rightarrow5s5p~^3P_2$ is investigated. It is found that no common magic wavelength could be identified for achieving triple magic trapping with the linearly polarized light. However, if the degree of ellipticity is between $0.358$ and $1$, the triple magic trapping can be achieved at 419.88 nm for the $5s^2 ~ ^1S_0\rightarrow 5s5p~^3P_2$ ($M_{i}=\pm2$) and $5s^2~^1S_0\rightarrow5s5p~^3P_{0}$ transitions.

\end{abstract}

\maketitle

\section{INTRODUCTION}

The development of state-of-the-art optical atomic clocks has been made significant progress thanks to advances in laser technologies such as optical frequency combs~\cite{udem2002, Cundiff2003}, narrow linewidth lasers~\cite{Young1999, Matei2017}, and frequency transmission technologies using optical fibers~\cite{riehle2017}. Currently, the highest precision of optical clocks is below $10^{-18}$~\cite{campbell2017, oelker2019, Brewer2019, zheng2022}. To evaluate the performance of optical clocks in experiments, a common approach is to compare two independent clocks to assess their stability, repeatability, and system uncertainty~\cite{bloom2014,Kobayashi2022,ushijima2015,zheng2022,Tyumenev_2016,Ohmae2020,kim2023}. Such comparisons were made with the same or different types of atoms or ions. 

 It is known that some ions or atoms, such as Yb~\cite{Safronova2018,Dzuba2018,Ishiyama2023,Tang2023}, Yb$^+$~\cite{Lange2021,Filzinger2023}, Al$^+$~\cite{bohman2023} and Sr~\cite{Trautmann2023}, possess two long-lived metastable states. Recently, it has been demonstrated that  transitions from these metastable states to ground states can be used as clock transitions with very narrow linewidths. This implies that two clock transitions can be measured simultaneously in the same atomic clock. By comparing these two clock transitions within a single atomic clock, it is possible to eliminate system effects and investigate the variation of the fine structure constant~\cite{Godun2014,Lange2021,Filzinger2023,bohman2023,Trautmann2023,Raizen2022}. In recent experiments, the optical frequency ratio of the $^2S_{1/2}(F=0)\rightarrow {^2F_{7/2}(F=3)}$ and  $^2S_{1/2}(F=0)\rightarrow {^2D_{3/2}(F=2)}$ transitions in $^{171}$Yb$^+$  was measured by Filzinger \emph{et al.}~\cite{Filzinger2023}, improving the existing limits on a linear temporal drift of the fine structure constant~\cite{Filzinger2023};  Bohman \emph{et al.} pointed out that the  ${^1S_0}\rightarrow {^3P_2}$ transition of the Al$^+$ ions is another clock transition and can be used to search for physics beyond the Standard Model~\cite{bohman2023}; Trantmann \emph{et al.}~\cite{Trautmann2023} have proposed triple magic trapping in Sr atoms, where the Stark shifts for the $5s^2~^1S_0$, $5s5p~^3P_0$, and $5s5p~^3P_2$ states are identical, to eliminate the Stark shifts of the two clock transitions.

Cd atoms have two valence electrons, and the $5s^2~^1S_0\rightarrow5s5p~^3P_0$ transition is an ultranarrow clock transition. Cd atoms are promising candidate for compact and transportable optical clocks~\cite{Ovsiannikov2007,Kaneda2016,Dzuba_2019,yamaguchi2019,Takamoto2022}. Yamaguchi~\emph{et al.}  achieved magic trapping at 419.88(14) nm~\cite{yamaguchi2019}, where the Stark shifts of the $5s^2~^1S_0$ and $5s5p~^3P_0$ states are equal~\cite{ye1999,katori1999}. In theory, there are some works~\cite{Dzuba_2019,Ye2008,Porsev2020,Sahoo2018,zhou2021} to calculate the black-body radiation (BBR) shifts, the magic wavelength of the $5s^2~^1S_0\rightarrow5s5p~^3P_0$ clock transition, and the multipolar polarizabilities and hyperpolarizabilities of the $5s^2~^1S_0$ and $5s5p~^3P_0$ states.
Similar to Sr atoms,  the $5s5p ^3P_2$ state of Cd is also a long-lived metastable state. As shown in Table~\ref{Tab1}, the lifetime of $5s5p~^3P_2$ is very close to that of the $5s5p~^3P_0$ state. Therefore, the $5s^2~^1S_0\rightarrow5s5p~^3P_2$ transition of Cd can also be considered as a second clock transition. Moreover, the static differential polarizability between $5s^2~^1S_0$ and $5s5p~^3P_2$ states is similar to that between $5s^2~^1S_0$ and $5s5p~^3P_0$ states. In addition, the BBR shifts experienced by both the $5s^2~^1S_0 \rightarrow 5s5p~^3P_0$ and $5s^2~^1S_0 \rightarrow 5s5p~^3P_2$ transitions are an order of magnitude smaller than those of Sr and Yb lattice clocks operating at the same room temperature~\cite{Porsev2006,Dzuba2019}.

\begin{table}
\caption[]	{The lifetimes (in second) of the $5s5p~^3P_2$ and $5s5p~^3P_0$ states of Cd~\cite{MISHRA2001}.}\label{Tab1}
\begin{ruledtabular}
	\begin{tabular}{ccc}
		Atom &  $\tau (5s5p~^3P_2)$ & $\tau (5s5p~^3P_0)$\\ 
		\midrule
		$^{111}$Cd & 19.0 & 17.9 \\
		$^{113}$Cd & 18.0 & 16.1 \\
	\end{tabular}
\end{ruledtabular}
\end{table}	

In this manuscript, we employ the relativistic configuration interaction plus second-order many-body perturbation theory (RCI+MBPT) method to calculate the dynamic polarizabilities of the $5s^2~^1S_0$ and $5s5p~^3P_{0,2}$ states in Cd. The magic wavelengths of the $5s^2~^1S_0\rightarrow5s5p~^3P_{0,2}$ transitions are confirmed based on the dynamic polarizabilities. We then investigate the conditions required for achieving triple magic trapping for the $5s^2~^1S_0\rightarrow5s5p~^3P_{0}$ and $5s^2~^1S_0\rightarrow5s5p~^3P_2$ transitions near the  magic wavelength of 419.88 nm. Throughout this paper, atomic units are used unless otherwise specified.

\begin{figure*}[tbh]	
\centering{
	\includegraphics[width=14.0cm, height=6cm]{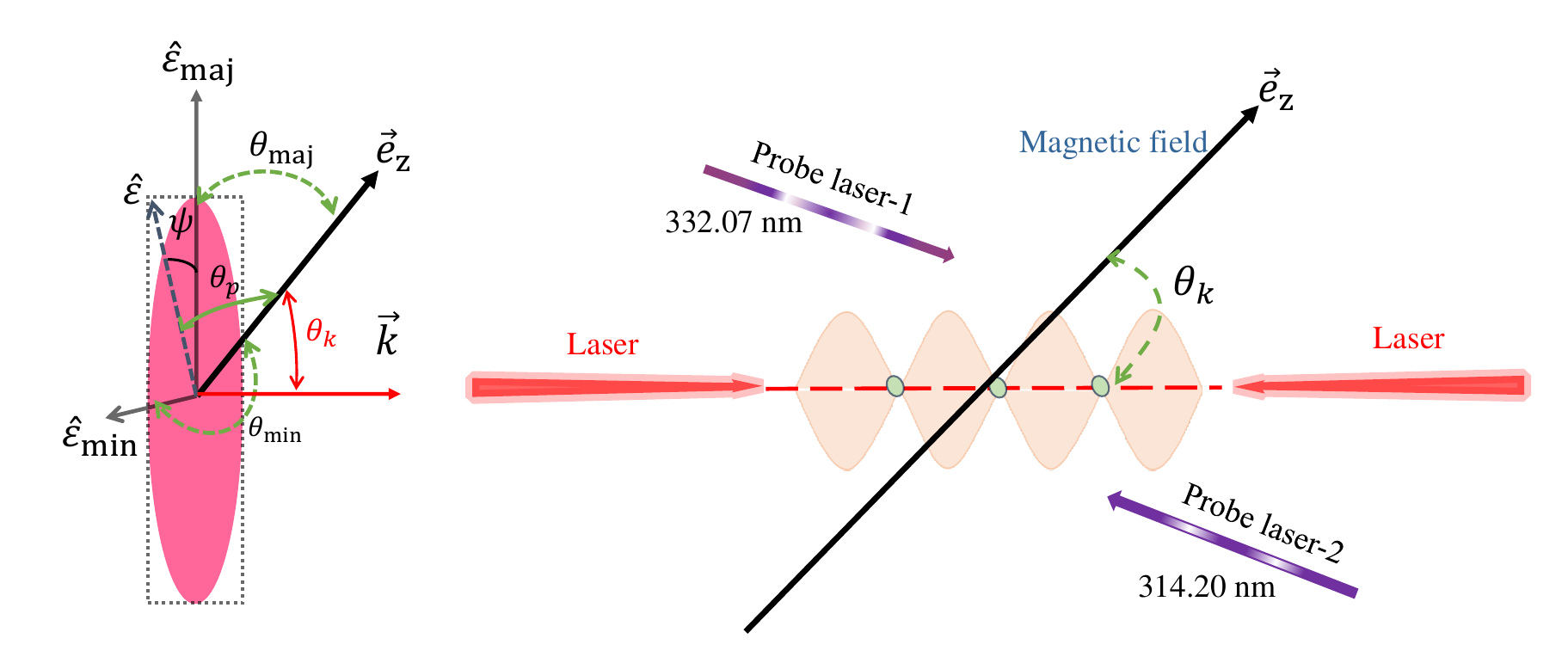}} 
\caption{\label{fig1} The schematic of the geometrical parameters of the electromagnetic plane wave and the 1D optical lattice trap. The elliptical area is swept by the electric field vector in one period. The unit vector $\hat{\varepsilon}_\text{maj}$ ($\hat{\varepsilon}_\text{min}$) is aligned with the semimajor (-minor) axis of the ellipse. $\vec{e}_\text{z}$ is the quantization axis, which determines the direction of the magnetic field in the experiment. $\vec{k}$ is the direction of the wave vector. $\theta_p$ is the angle between the polarization vector $\hat{\varepsilon}$ and the quantization axis $\vec{e}_\text{z}$, $\theta_k$ is the angle between $\vec{e}_\text{z}$ and $\vec{k}$. The $\hat{\varepsilon}_\text{maj}$, $\hat{\varepsilon}_\text{min}$, and $\vec{k}$ are orthogonal to each other. $\theta_\text{maj}$ ($\theta_\text{min}$) is the angle between $\hat{\varepsilon}_\text{maj}$ ($\hat{\varepsilon}_\text{min}$) and $\vec{e}_\text{z}$. $\psi$ is directly related to the degree of circular polarization of the electromagnetic plane wave. Probe laser-1 and Probe laser-2  represent two probe lasers with wavelengths of 332.07 and 314.20 nm of the two clock transitions $5s^2~^1S_0\rightarrow 5s5p~^3P_{0}$ and $5s^2~^1S_0\rightarrow 5s5p~^3P_2$, respectively.}
\end{figure*} 
\begin{table}
  \caption []{Pseudospectral oscillator strength distribution for Cd$^{2+}$ ions. The transition energies $\Delta E_{n\rightarrow i}$ are given in a.u..}\label{Tab2I}
  \begin{ruledtabular}
  	\begin{tabular}{crr}	
  		$n$ & \multicolumn{1}{c}{$\Delta E_{n\rightarrow i}$} & \multicolumn{1}{c}{$f_n$} \\
  		\midrule
  		1 & 961625.7331 & 2.0 \\
  		2 & 20777.9141  & 2.0 \\
  		3 & 565.0139  & 2.0 \\
  		4 & 0.6154  & 2.0 \\
  		5 & 352.5699  & 6.0 \\
  		6 & 132.0840  & 6.0 \\ 
  		7 & 5.5682  & 6.0 \\
  		8 & 67.3807  & 10.0 \\
  		9 & 23.0199  & 10.0 \\  		
		\end{tabular}
\end{ruledtabular}				
\end{table}
\section{Theory}\label{Sec.II}

\subsection{The expression of the polarizabilities}\label{Sec.II-B}

\begin{table}
	\caption []{Comparison of energy levels (in cm$^{-1}$) for some low-lying states. All the energies are given relative to the ground state of the Cd$^{2+}$ core. The relative differences between the present energy and the NIST energy~\cite{NIST_ASD} are listed as Diff.}\label{Tab2}
	\begin{ruledtabular}
		\begin{tabular}{cccc}
			{State} & {Present} & {NIST~\cite{NIST_ASD}} &  Diff. \\			
			\midrule
			$5s^2~^{1}S_{0}$ & $-$208818.3& $-$208914.8  & 0.05\%  \\
			$5s5p~^{3}P^o_{0}$ & $-$178829.2& $-$178800.8 & 0.02\%  \\
			$5s5p~^{3}P^o_{1}$ & $-$178291.1& $-$178258.7 & 0.02\%  \\
			$5s5p~^{3}P^o_{2}$  & $-$177138.6  & $-$177087.8 & 0.03\%  \\
			$5s5p~^{1}P^o_{1}$ & $-$166007.0& $-$165222.4 & 0.47\%  \\
			$5s6s~^{3}S_{1}$ & $-$157917.5& $-$157430.8 & 0.31\%  \\
			$5s6s~^{1}S_{0}$ & $-$156035.7& $-$155604.7  & 0.28\%  \\
			$5s6p~^{3}P^o_{0}$ & $-$151075.4& $-$150523.9 & 0.37\% \\
			$5s6p~^{3}P^o_{1}$ & $-$151004.8& $-$150453.2 & 0.37\%  \\
			$5s6p~^{3}P^o_{2}$  & $-$150829.1  & $-$150279.1 & 0.37\%  \\
			$5s5d~^{1}D_{2} $  & $-$150217.9 & $-$149695.0  & 0.35\% \\
			$5s5d~^{3}D_{1}$ & $-$150019.9& $-$149429.0 & 0.40\% \\
			$5s5d~^{3}D_{2} $  & $-$150008.7 & $-$149416.9  & 0.40\%  \\
			$5s6p~^{1}P^o_{1}$ & $-$149642.4& $-$149007.5 & 0.43\%  \\
			$5s7s~^{3}S_{1}$ & $-$146945.8& $-$146351.3 & 0.41\%  \\
			$5s7s~^{1}S_{0}$ & $-$146398.1& $-$145827.9  & 0.39\% \\
			$5s7p~^{3}P^o_{0}$ & $-$144522.3& $-$143918.9 & 0.42\% \\
			$5s7p~^{3}P^o_{1}$ & $-$144496.1& $-$143889.3 & 0.42\%  \\
			$5s7p~^{3}P^o_{2}$  & $-$144427.7  & $-$143821.1 & 0.42\%  \\
			$5s6d~^{1}D_{2} $  & $-$144364.5 & $-$143780.0  & 0.41\%  \\
			$5s6d~^{3}D_{1}$ & $-$144184.8& $-$143561.4 & 0.43\% \\
			$5s6d~^{3}D_{2} $  & $-$144179.6 & $-$143555.9  & 0.43\% \\
			$5s7p~^{1}P^o_{1}$ & $-$144044.6 & $-$143413.4 & 0.44\% \\
			
			$5s4f~^{3}F^o_{2}$  & $-$143971.2  & $-$143328.7 & 0.45\% \\
			$5s8s~^{3}S_{1}$ & $-$142854.1& $-$142232.7 & 0.44\%  \\
			$5s8s~^{1}S_{0}$ & $-$142618.6& $-$142009.1  & 0.43\%  \\
			$5s8p~^{3}P^o_{0}$ & $-$141710.1& $-$141085.1 & 0.44\% \\
			$5s7d~^{1}D_{2} $  & $-$141686.2 & $-$141076.4  & 0.43\%  \\
			$5s8p~^{3}P^o_{2}$  & $-$141663.3  & $-$141039.6 & 0.44\% \\
			$5s7d~^{3}D_{1}$ & $-$141557.7& $-$140925.0 & 0.45\%  \\
			$5s7d~^{3}D_{2} $  & $-$141554.9 & $-$140922.1  & 0.45\%  \\
			$5s5f~^{3}F^o_{2}$  & $-$141461.7  & $-$140821.1 & 0.45\% \\
			$5s9s~^{1}S_{0}$ & $-$140740.4& $-$140116.0 & 0.45\%  \\		
			$5s9p~^{3}P^o_{0}$ & $-$140232.2& $-$139600.7 & 0.45\% \\
			
		\end{tabular}
	\end{ruledtabular}				
\end{table}
\begin{table*}
\caption[]{Comparison of some reduced $E1$ matrix elements (in a.u.) for principal transitions of Cd. The numbers in the parentheses are uncertainties.}\label{Tab3}
\begin{ruledtabular}
	\begin{tabular}{lccccc}			
		{Transition} & {Present}& {CI+All~\cite{yamaguchi2019}}& {CI+MBPT}& {DFCP~\cite{zhou2021}} & {Expt${^\textrm{c}}$.~\cite{NIST_ASD}}\\
		\midrule
		
		$5s^2~^1S_{0}\rightarrow5s5p~^1P_{1}$ & 3.479 & 3.440 & 3.426~\cite{yamaguchi2019} & 3.4787${^\textrm{a}}$& 3.01\\
		&   &  & 3.435~\cite{Dzuba2019} & 3.4292${^\textrm{b}}$\\
		
		$5s^2~^1S_{0}\rightarrow5s5p~^3P_{1}$ & 0.167 && 0.158~\cite{Dzuba2019} & &  0.14\\
		$5s^2~^1S_{0}\rightarrow5s6p~^1P_{1}$ & 0.670 & 0.689 & 0.675~\cite{yamaguchi2019} & 0.5957${^\textrm{a}}$\\
		&   &  & & 0.6552${^\textrm{b}}$ \\
		
		$5s5p~^3P_{0}\rightarrow5s6s~^3S_{1}$ & 1.493 & 1.491 & 1.502~\cite{yamaguchi2019} & 1.6085${^\textrm{a}}$ & 1.42\\
		&   &  & 1.486~\cite{Dzuba2019} & 1.5619${^\textrm{b}}$\\
		$5s5p~^3P_{0}\rightarrow5s5d~^3D_{1}$ & 2.325 & 2.318 & 2.306~\cite{yamaguchi2019} & 2.4537${^\textrm{a}}$ & 2.12\\
		&   &  & 2.222~\cite{Dzuba2019} & 2.3667${^\textrm{b}}$\\	
		$5s5p~^3P_{0}\rightarrow5s7s~^3S_{1}$ & 0.433 & 0.433 & 0.432~\cite{yamaguchi2019} & 0.4479${^\textrm{a}}$ & \\ 	
		&   &  & & 0.4445${^\textrm{b}}$\\
		$5s5p~^3P_{0}\rightarrow5s6d~^3D_{1}$ & 1.066 & 1.061 & 1.062~\cite{yamaguchi2019} & 1.0998${^\textrm{a}}$ & 0.97\\
		&   & && 1.0778$^\textrm{b}$\\
		$5s5p~^3P_{1}\rightarrow5s5d~^3D_{2}$ & 3.543 &&&& 3.51\\	
		$5s5p~^3P_{1}\rightarrow5s6s~^3S_{1}$ & 2.642 &&&& 2.58\\
		$5s5p~^3P_{1}\rightarrow5s5d~^3D_{1}$ & 2.044 &&&& 2.03\\
		$5s5p~^3P_{2}\rightarrow5s5d~^3D_{3}$ & 5.045 &&&& 4.63\\
		$5s5p~^3P_{2}\rightarrow5s6s~^3S_{1}$ & 3.592 &&&& 3.30\\
		$5s5p~^3P_{2}\rightarrow5s5d~^3D_{2}$ & 2.128 &&&& 2.03\\
		$5s5p~^1P_{1}\rightarrow5s5d~^1D_{2}$ & 5.427 &&&& 6.23\\	
		$5s5p~^1P_{1}\rightarrow5s6s~^1S_{0}$ & 3.923 &&&& \\
	\end{tabular}
\end{ruledtabular}	
\begin{tablenotes}
	\footnotesize
	\item{${^\textrm{a}}$}represents the values obtained without including the high-order one-body and two-body core-polarization potentials.
	\item{${^\textrm{b}}$}represents the values obtained with the high-order one-body and two-body core-polarization potentials.
	\item{${^\textrm{c}}$}The values of the experimental matrix elements are obtained from the oscillator strengths.			
\end{tablenotes}
\end{table*}

When an ion or atom exposed to a laser field with the laser frequency $\omega$, degree of ellipticity $\mathcal{A}$, direction of the wave vector $\vec{k}$, as shown in Figure~\ref{fig1}, the energy shift due to the Stark effect can be written as~\cite{mitroy2010}
\begin{equation}\label{eq3}
 \Delta E_i =-\frac{1}{2}\alpha_i(\omega)F^2+\cdots,
\end{equation}
where $\alpha_i(\omega)$ is the dynamic dipole polarizability of the quantum state $i$, and $F$ is a measure of the strength of the ac electromagnetic field. The dipole polarizability can be calculated using the sum-over-states method, and
the general expression of dynamic dipole polarizability can be written as~\cite{manakov1986,beloy2009,le2013,mitroy2010}
\begin{eqnarray}\label{eq4}
\alpha_i(\omega)=&&\alpha^S_i(\omega)+\mathcal{A}\text{cos}\theta_k \frac{M_i}{2J_i} \alpha^V_i(\omega)
\nonumber \\
&&+\frac{3\text{cos}^2\theta_p-1}{2}\frac{3M^2_i-J_i(J_i+1)}{J_i(2J_i-1)}\alpha^T_i(\omega),
\end{eqnarray}
where $M_{i}$ is the component of the total angular momentum $J_i$.  $\theta_k$ is the angle between $\vec{e}_\text{z}$ and $\vec{k}$,  $\vec{e}_\text{z}$ is the quantization axis (the direction of static magnetic field), and  $\text{cos}\theta_k=\vec{k}\cdot\vec{e}_\text{z}$. 
Geometrically, $\theta_p$ is related to the $\theta_{\text{maj}}$ and $\theta_{\text{min}}$, it can be written as
\begin{eqnarray}\label{eq5}
	\text{cos}^2\theta_p&&=\text{cos}^2\psi \text{cos}^2\theta_\text{maj}+\text{sin}^2\psi \text{cos}^2\theta_\text{min}
	\nonumber\\
	&&=\text{sin}^2\psi \text{sin}^2\theta_{k}+\text{cos}(2\psi)\text{cos}^2\theta_\text{maj},
\end{eqnarray}
here, $\theta_{\text{maj}}$ ($\theta_{\text{min}}$) is the angle between the major (minor) axis of the ellipse and the $\vec{e}_\text{z}$ axis.
For linearly polarized light, $\theta_p$ is the angle between the laser polarization vector $\hat{\varepsilon}$ and the $\vec{e}_\text{z}$ axis.
The degree of ellipticity $\mathcal{A}$ is directly related to the angle $\psi$ $(\left|\psi\right|\leq 45^\circ)$, 
\begin{eqnarray}\label{eq6}
	\mathcal{A}=\text{sin}2\psi.
\end{eqnarray}
$\mathcal{A}=0$ is linearly polarized light, $\mathcal{A}=+1$ and $-1$ are the right-handed and left-handed circularly polarized light, respectively. $\alpha^S_i(\omega)$, $\alpha^V_i(\omega)$, and $\alpha^T_i(\omega)$ are the scalar, vector, and tensor polarizabilities, they can be expressed as
\begin{eqnarray}\label{eq7}
\alpha^S_i(\omega)=\frac{2}{3(2J_i+1)}
\sum_{n}\frac{\varepsilon_\text{in}{\langle \gamma_n J_n\lVert d\rVert \gamma_i J_i\rangle}^2}{\varepsilon^2_\text{in}-\omega^2},
\end{eqnarray}

\begin{eqnarray}\label{eq8}
&&\alpha^V_i(\omega)=-2\sqrt{\frac{6J_i}{(J_i+1)(2J_i+1)}}     
\nonumber \\
&&\times\sum_{n}{(-1)^{J_i+J_n}
	\left\{
	\begin{array}{ccc} 
		1 & 1 & 1 \\
		J_i & J_i & J_n\\
	\end{array}
	\right\}
	\frac{\omega\langle \gamma_n J_n \lVert d\rVert \gamma_i J_i\rangle^2}{\varepsilon^2_\text{in}-\omega^2}},
\end{eqnarray}
and
\begin{align}\label{eq9}
&\alpha^T_i(\omega)= 4\left(\frac{5J_i(2J_i-1)}{6(J_i+1)(2J_i+1)(2J_i+3)}\right)^{1/2}
\nonumber \\
& \times\sum_{n}(-1)^{J_i+J_n}
\left\{
\begin{array}{ccc}
	J_i & 1 & J_n \\ 
	1 & J_i & 2   \\ 
\end{array}
\right\} 
\frac{{\langle \gamma_n J_n\lVert d\rVert \gamma_i J_i\rangle}^2\varepsilon^2_\text{in}}{\varepsilon^2_\text{in}-\omega^2},
\end{align}
where  $\varepsilon_\text{in}=E_n-E_i$ is excitation energy, 
$\langle \gamma_n J_n \lVert d\rVert \gamma_i J_i\rangle$ is the reduced $E1$ transition matrix element. If $\omega=0$, the dynamic polarizabilities in Eqs. (\ref{eq7}) and (\ref{eq9}) are reduced to the static polarizabilities, and the static vector polarizability equals to zero. 

The dynamic polarizability of the Cd$^{2+}$ core is calculated using a pseudospectral oscillator strength distribution method, similar to that used in Refs.~\cite{Mitroy2003,jiang2013,Margoliash1978,Ashok1985}. 
Table~\ref{Tab2I} lists the pseudospectral oscillator strength distribution
for Cd$^{2+}$, where the pseudo-oscillator strength $f_n$ is equal to the number of electrons in the shell. The excitation energy is set by adding a constant to the Koopman energies and adjusting the constant until the core polarizability, calculated from the oscillator strength sum rule, matches the known core polarizability 4.97 a.u.~\cite{JOHNSON1983,Aymar1996,Mitroy_2010}.

\subsection{RCI+MBPT Method}

The wave functions and energy levels of Cd are obtained by the RCI + MBPT calculation.  In this method, the many-electron atomic or ionic system is divided into a frozen core part and valence electron parts. The initial step is to perform a self-consistent Dirac-Fock (DF) calculation to obtain all the single-particle orbitals. These orbitals are then used to build the configuration space. In this step, we use the no-pair Dirac Hamiltonian and treat both the Coulomb and Breit interactions on an equal footing, similar to Refs.~\cite{tang2013pra,tang2014cpb,tang2017pra}.
The next step involves carrying out second-order many-body perturbation and configuration interaction calculations to account for the core-valence and valence-valence correlations.

The effective interaction equation for divalent atomic system can be expressed as
\begin{equation}
	\label{eq1}
	\left\{\sum^{2}_{i}{\left[H_{\rm DF}(r_{i})+\Sigma_{1}(r_{i})\right]}+(\frac{1}{r_{12}}+\Sigma_{2})\right\}|\gamma J\rangle=E|\gamma J\rangle,
\end{equation}
where $H_{\rm DF}$ and $\frac{1}{r_{12}}$ denote the DF Hamiltonian and electron-electron Coulomb interaction, respectively. 
$\Sigma_{1}$ is the one-body correlation potential, which describes the correlation interaction between a valence electron and the core. The $\Sigma_{2}$ represents the two-body correlation potential, which describes the screening of the Coulomb interaction between valence electrons and the core electrons. The matrix elements of the one-body and two-body correlation potentials have been given in Ref.~\cite{Safronova2009pra}.
To account for the correlation effects beyond second-order, we introduce the rescaling parameter $\rho_{\kappa}$ and substitute the one-body correlation potentials $\Sigma_{1}$ with $\rho_{\kappa}\Sigma_{1}$ in practice. The rescaling parameter $\rho_{\kappa}$ is tuned to reproduce the experimental energy of the lowest state for each angular quantum number $\kappa$ of a monovalent atomic system, which is similar to the Dirac-Fock plus core polarization (DFCP) method~\cite{tang2013pra}. The rescaling parameter can accelerate the convergence of RCI+MBPT method.

For the transition matrix element calculation, it is necessary to consider the core polarization correction. In present work, we take the core polarization correction into account by using the Random Phase Approximation (RPA)~\cite{Dzuba1998JETP,Savukov2000pra}. The further details on RCI+MBPT method can be found in Refs.~\cite{Dzuba1996pra,Dzuba1998pra,Porsev2001pra,Savukov2002pra,Safronova2009pra,zhang2023}.

In the DF calculation, the large and small components of the Dirac wave functions are expanded using 50 B-spline bases of order $k=13$ and the box size $R_\text{max}=240$. The partial waves are limited to $\ell_\text{max} = 5$ and the lowest 40 orbital sets of each partial wave are used to construct the configuration space. In the second order many-body perturbation calculations, the summation is carried out over the entire basis set. The rescaling parameters are $\rho_{-1}=0.902$, $\rho_{1}=0.971$, $\rho_{-2}=0.982$, $\rho_{2}=1.028$, $\rho_{-3}=1.037$, and $\rho_{\rm others}=1.0$, respectively.

\section{results and discussion} \label{Sec.III}

\subsection{Energy levels, reduced $E1$ matrix elements and static polarizabilities}

Table~\ref{Tab2} lists the presently calculated energy levels and compares them with the National Institute of Standards and Technology (NIST) tabulations~\cite{NIST_ASD}. Our RCI + MBPT results are in excellent agreement with the NIST results, especially for the low-lying states such as the $5s^2~^1S_{0}$, $5s5p~^3P_0$, $^3P_1$, and $^3P_2$ states, with differences of only 0.05\%, 0.02\%, 0.02\%, and 0.03\%, respectively. For other excited states, the largest discrepancy does not exceed 0.5\%. 

 \begin{table*}
 	\caption[]	{Comparison of the static dipole polarizabilities (a.u.) of the low-lying states of Cd with available experimental and theoretical results. Numbers in parentheses represent the uncertainties in the last digits. The polarizability of the core in present work is $\alpha_{\text{core}}=4.97$ a.u.~\cite{JOHNSON1983,Aymar1996,Mitroy_2010}.}\label{Tab4}
 	\begin{ruledtabular}
 		\begin{tabular}{cllllll}
 			& \multicolumn{3}{c}{$\alpha^S$} & \multicolumn{3}{c}{$\alpha^T$} \\
 			\midrule
 			& \multicolumn{1}{c}{Present} & \multicolumn{1}{c}{Refs.} & \multicolumn{1}{c}{Expt.} & \multicolumn{1}{c}{Present} & \multicolumn{1}{c}{Refs.} & \multicolumn{1}{c}{Expt.}\\
 			\cmidrule(r){1-4} \cmidrule(r){5-7}
$5s^2~^1S_0$ & 47.4(1.5) &  46.53 CI+All~\cite{yamaguchi2019}   & 49.65(165)~\cite{Goebel1995}\\ 
 		              	&& 46.52 CI+MBPT~\cite{Dzuba2019} & 47.5(2.0)~\cite{Uwe2022}\\
 		        	&& 46.02(50) RCC~\cite{Sahoo2018}\\
 			       && 46.9 DHF~\cite{Roos2005}\\
 		         	&& 46(2) DFCP~\cite{zhou2021}\\
 			       && 45.92(10) RCCSD~\cite{Guo2021}\\
 			       && 44.63 CPMP~\cite{Ye2008} \\
$5s5p~^3P_0$ & 77.3(2.3) & 76.2 CI+All~\cite{yamaguchi2019}&    \\ 
 			&& 78(6) DFCP~\cite{zhou2021}\\ 
 			&& 75.31 CI+MBPT~\cite{Dzuba2019}\\
 			&& 75.03(42) RCCSD~\cite{Guo2021}\\
 			&&75.29 CPMP~\cite{Ye2008} \\
 
$5s5p~^3P_1$ & 80.2(2.2)  & 77.87(40) RCCSD~\cite{Guo2021} & & 7.74(33) & 7.30(7) RCCSD~\cite{Guo2021} &7.11(32)~\cite{rinkleff1979}\\
 			 &&&&& 6.30 Model-potential~\cite{robinson1969} & 5.35(16)~\cite{legowski1995} \\
 			 &&&&&& 6.91~\cite{bonch1968}\\
 			 &&&&&& 6.83(28)~\cite{Khadjavi1968}\\
$5s5p~^3P_2$ & 87.7(1.8) & 84.82(34) RCCSD~\cite{Guo2021} &&{$-$19.15(84)} & {$-$18.05(13) RCCSD~\cite{Guo2021}}\\
$5s5p~^1P_1$ &218.0(6.2) & 225.18(2.59) RCCSD~\cite{Guo2021}&& {$-$84.9(2.6)} & $-$90.99(1.75) RCCSD~\cite{Guo2021}\\

\end{tabular}
\end{ruledtabular}
\end{table*}

Table~\ref{Tab3} presents the reduced matrix elements for some of the main transitions and compares them with other available results. The present results for the transitions from the $5s^2~^1S_{0}$ state agree well with those of the configuration interaction plus an all-order linearized coupled-cluster (CI+All) method~\cite{yamaguchi2019}, the configuration interaction plus many-body perturbation theory (CI+MBPT)~\cite{yamaguchi2019,Dzuba2019}, and the DFCP method~\cite{zhou2021}, the differences are about 1\%, 1\% and 2\%, respectively. For the transitions from the $5s5p~^3P_{0}$ state, the differences between the present results with these three methods are about 0.3\%, 0.5\%, and 2.5\%, respectively. Regarding the transitions from  the $5s5p~^3P_1$, $5s5p~^3P_2$ and $5s5p~^1P_1$ states, there are no other theoretical results available for comparison. However, all of the present results are in agreement with those of NIST~\cite{NIST_ASD}. 
\begin{figure}[tbh]	
	\centering{
		\includegraphics[width=8.5cm, height=7.0cm]{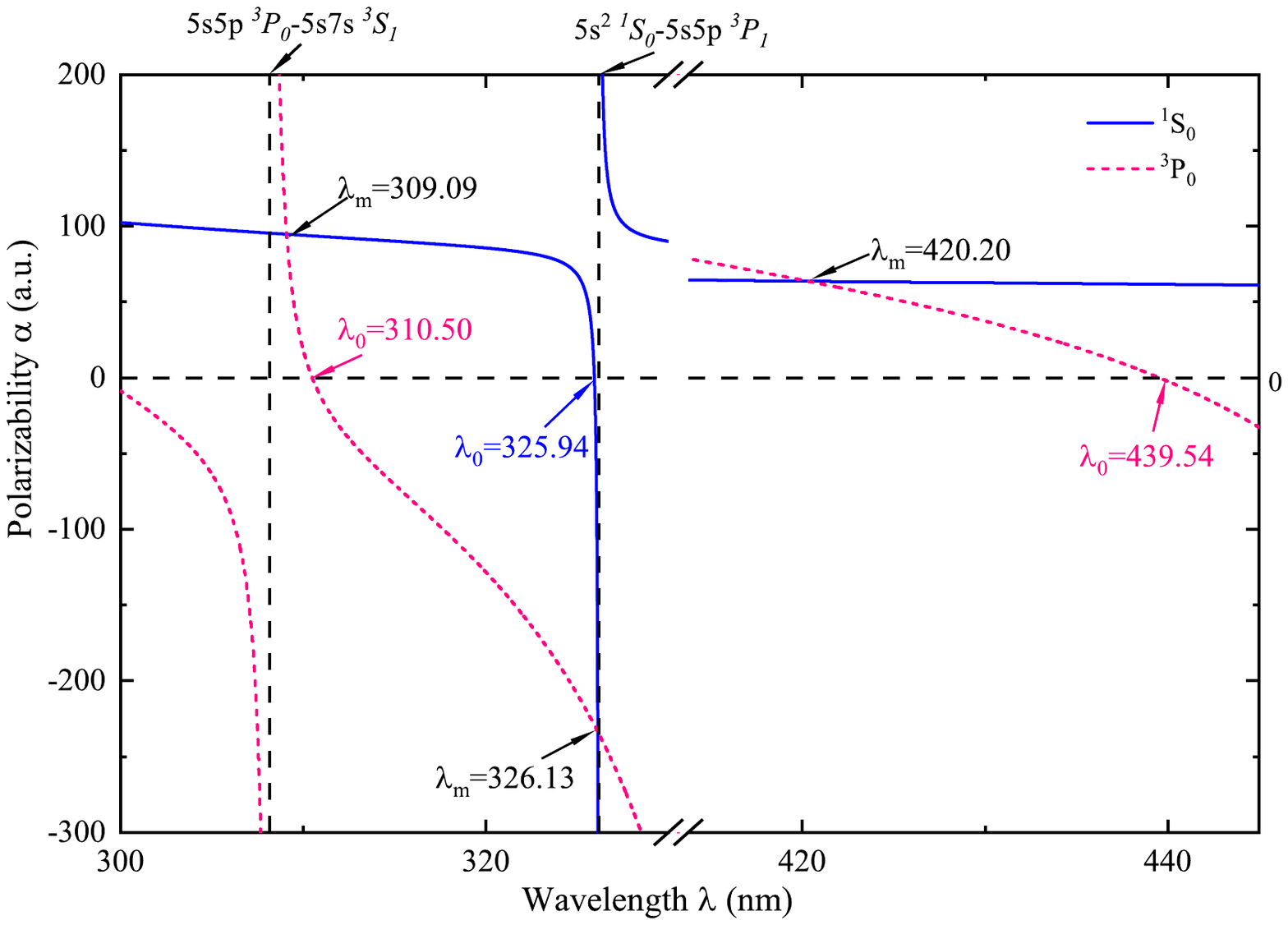}} 
	\caption{\label{fig2} The dynamical polarizabilities of $5s^2~^1S_{0}$ and $5s5p~^3P_0$ states  in the $300-500$ nm wavelength range. The vertical dashed lines indicate the positions of the resonant transitions. The magic-zero wavelengths ($\lambda_0$) are determined by locating points where either $5s5p~^3P_0$ or $5s^2~^1S_{0}$ polarizabilities are equal to zero and the magic wavelengths ($\lambda_\text{m}$) are determined by locating points where the $5s5p~^3P_0$ and $5s^2~^1S_{0}$ polarizabilities are equal to each other. They are all identified by arrows.}
\end{figure}

Table~\ref{Tab4} lists the results of the static dipole polarizabilities of the $5s^2~^1S_{0}$, $5s5p~^3P_{0,1,2}$ and $5s5p~^1P_{1}$ states. In the present calculations, the experimental transition energies are used for the low-lying excited states. The numbers of the intermediate states are $N_{J^P=1^+}=10457$, $N_{J^P=1^-}=10140$, $N_{J^P=2^+}=14390$, $N_{J^P=3^+}=15077$ , where $+$ and $-$ represent even and odd parity, respectively. For the $5s^2~^1S_{0}$ state, the present static polarizability of 47.4(1.5) a.u. is in excellent agreement with the experimental value of 47.5(2.0) a.u.~\cite{Uwe2022}. The differences between the present result and the CI+All and CI+MBPT results are about 0.9 a.u. These differences are mainly due to the difference in the reduced matrix elements, as shown in Table~\ref{Tab3}. If we replace the present reduced matrix element of the $5s^2~^1S_{0}\rightarrow5s5p~^1P_1$ transition, which dominantly contributes to the polarizability of the $5s^2~^1S_{0}$ state, with the CI+All value of 3.440~\cite{yamaguchi2019} or CI+MBPT value of 3.435 a.u.~\cite{Dzuba2019},  while the corresponding value becomes 46.17 and 46.38 a.u. respectively. The differences between the present result and the results of relativistic coupled-cluster theory (RCC)~\cite{Sahoo2018},  DFCP method~\cite{zhou2021}, and relativistic coupled-cluster single-double excitations approach (RCCSD)~\cite{Guo2021} are about 1.4 a.u. These differences are mainly caused by the difference in reduced matrix elements and transition energies. The uncertainties given in this table are determined by introducing 2\% changes in the dominant matrix elements, since most of the present results agree with other results within 2\%.

For the $5s5p~^3P_0$ state, the differences between the present calculations and the other theoretical results are within 3\%.  As for $5s5p~^3P_1$, $5s5p~^3P_2$, and $5s5p~^1P_{1}$ states, it is found that the present results are in agreement with the results of the RCCSD calculations~\cite{Guo2021}. The tensor polarizability of the $5s5p~^3P_{1}$ state, 7.74(22) a.u., is in good agreement with the experimental value of 7.11(32) a.u.~\cite{rinkleff1979}.
  
 It should be noted that the differential polarizabilities $\alpha(5s5p~^3P_0)-\alpha(5s^2~^1S_0)$ and $\alpha(5s5p~^3P_2~M_i=\pm2)-\alpha(5s^2~^1S_0)$ are 29.9(2.8) and 21.1(2.5) a.u., respectively, as shown in Table~\ref{Tab5}. The difference between these two values is small. This means that at the same temperature the BBR shift of the $5s^2~^1S_0\rightarrow5s5p~^3P_2~M_i=\pm2$ transition is similar to that of the $5s^2~^1S_0\rightarrow5s5p~^3P_0$ transition.
 
  \subsection{The magic wavelengths for linearly polarized light}

 For linearly polarized light, that is $\psi=0$ and $\mathcal{A}=0$, Eq.~(\ref{eq4}) can be simplified as: 
 \begin{eqnarray}\label{eq10}
 	\alpha_i(\omega)=\alpha^S_i(\omega)+\frac{3\text{cos}^2\theta_p-1}{2}\frac{3M^2_{i}-J_i(J_i+1)}{J_i(2J_i-1)}\alpha^T_i(\omega),
 	\nonumber \\
 \end{eqnarray}	
where $\theta_p$ satisfies $0\leq \text{cos}^2\theta_p\leq1$. For states with $J_i \geq 1$, the polarizabilities include the scalar and tensor components, which are associated with $M_{i}$ and $\theta_p$.  For $J_i=0$ states, however, the polarizabilities are determined only by the scalar component. 

\begin{table}
	\caption[]	{Comparison of the differential polarizabilities    $\Delta\alpha({^3P_0}-{^1S_0})$ and $\Delta\alpha({^3P_2}-{^1S_0})$. The total polarizability $\alpha(5s5p~{^3P_2})$ is calculated by letting $\alpha^V_i=0$ and $\theta_p=0^\circ$ in Eq.~(\ref{eq4}).}\label{Tab5}
	\begin{ruledtabular}
		\begin{tabular}{cll}
			$\Delta\alpha$ & \multicolumn{1}{c}{Present} & \multicolumn{1}{c}{Refs.}  \\
			\midrule
			$\Delta\alpha({^3P_0}-{^1S_0})$ & 29.9(2.8) & 29.67 CI+All~\cite{yamaguchi2019}\\
			&& 28.79 CI+MBPT~\cite{Dzuba2019}\\
			&& 29.11(43) RCCSD~\cite{Guo2021}\\
			&& 30.66 CPMP~\cite{Ye2008}\\
			$\Delta\alpha({^3P_2}-{^1S_0})$\\
			$M_i=\pm2$ & 21.1(2.5)  & 20.85(38) RCCSD~\cite{Guo2021} \\ 
			$M_i=\pm1$ & 49.9(2.5)  & 47.93(38) RCCSD~\cite{Guo2021} \\ 
			$M_i=0$ & 59.4(2.5) & 56.95(38) RCCSD~\cite{Guo2021} \\ 
		\end{tabular}
	\end{ruledtabular}
\end{table}	
\begin{figure}[tbh]	
	\centering{
		\includegraphics[width=8 cm, height=17cm]{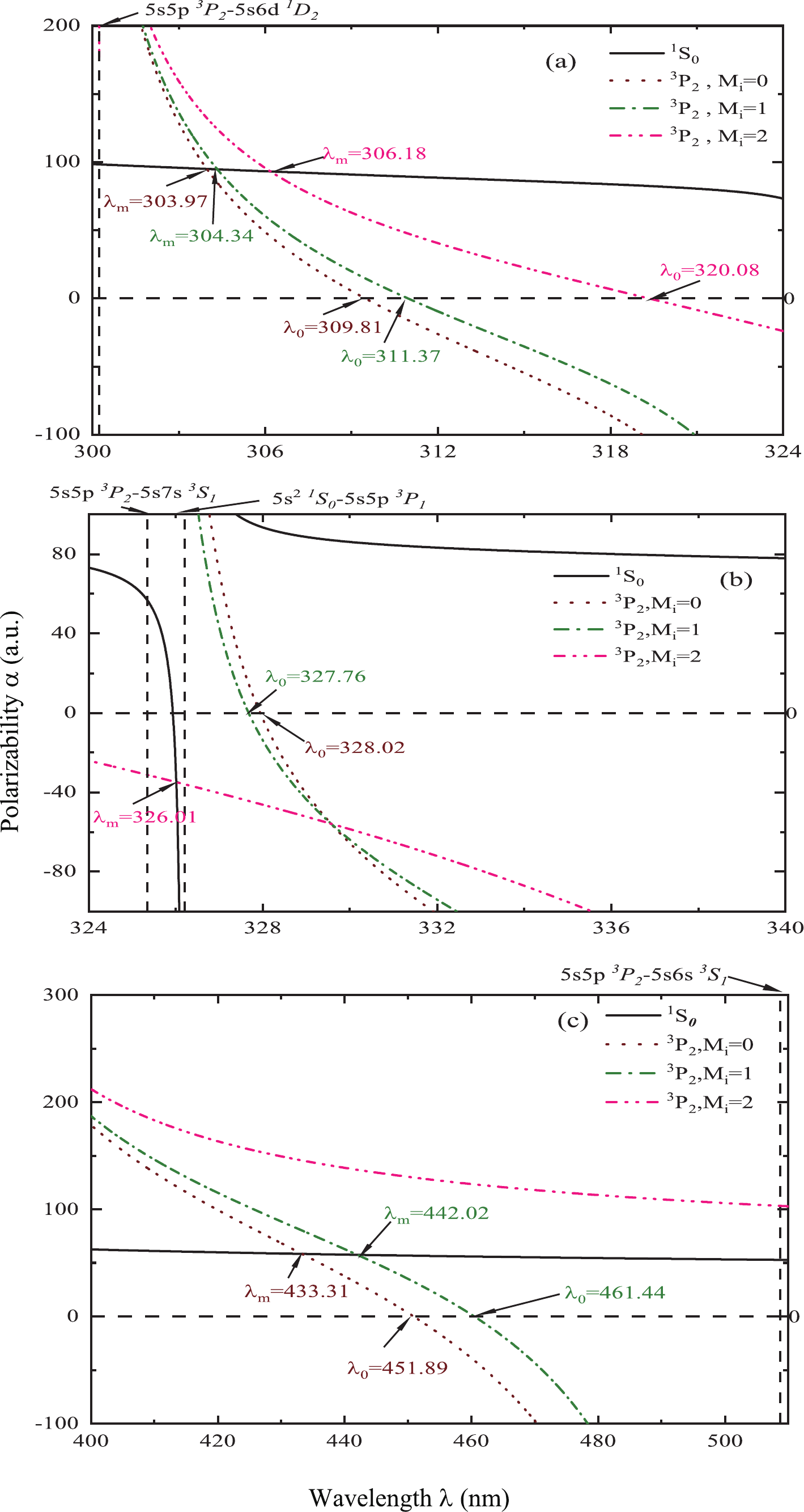} }
	\caption{\label{fig3} The dynamical polarizabilities of $^1S_0$ and $^3P_2$ states. The vertical dashed lines indicate the positions of the resonant transitions and are given on the top of the figures. The magic wavelengths ($\lambda_\text{m}$) are determined by locating points where the $5s5p~^3P_2$ and $5s^2~^1S_0$ polarizabilities are equal to each other, and the magic-zero wavelength ($\lambda_0$) of a state is determined when the polarizability is zero. They are all identified by arrows.}
\end{figure} 

Figure~\ref{fig2} shows the dynamic polarizabilities of the $5s^2~^1S_0$ and $5s5p~^3P_0$ states. Three magic wavelengths for the $5s^2~^1S_0\rightarrow 5s5p~^3P_0$ transition are found and identified by arrows. The longest magic wavelength obtained from our results is 420.20(57) nm, which agrees well with the experimental measurement of 419.88(14) nm~\cite{yamaguchi2019}, and theoretical calculations of 420.1(7) nm using the CI+All approach~\cite{yamaguchi2019} and 420 nm calculated using the CPMP approach~\cite{Ye2008}. The number in parentheses represents the uncertainty of the magic wavelength. The uncertainty was primarily caused by the uncertainty of the dominant matrix elements. In the present work, we change the dominant matrix elements by 2\% separately to calculate the change of the magic wavelength. Subsequently, the uncertainty of the magic wavelength is determined as the root-mean-square of each change. There are no comparable theoretical or experimental results for the other two magic wavelengths, 326.13(1) and 309.09(2) nm. These two magic wavelengths are very difficult to trap Cd atoms, because they are very close to the resonance wavelengths of the $5s^2~^1S_0\rightarrow5s5p~^3P_1$ and $5s5p~^3P_0\rightarrow5s7s~^3S_1$ transitions, which are 326.20 and 308.17 nm, respectively.

Figure~\ref{fig3} shows the dynamic polarizabilities of the $5s^2~^1S_0$ and each of the magnetic sublevels of the $5s5p~^3P_2$ states for the case $\theta_p=0^{\circ}$. The magic wavelengths, denoted by arrows, are listed in Table~\ref{Tab6}. It can be seen that there are two magic wavelengths for each of the magnetic sublevel transitions of the $5s^2~^1S_0\rightarrow 5s5p~^3P_2~\left|M_{i}\right|=0,1,2$. These magic wavelengths lie within two resonance intervals. The magic wavelengths near 300 nm lie between the resonance wavelengths of the $5s5p~^3P_2\rightarrow5s7s~^3S_1$ and $5s5p~^3P_2\rightarrow5s6d~^1D_2$ transitions. The longest magic wavelengths for each of these transitions lie between the resonance wavelengths of the $5s5p~^3P_2 \rightarrow5s6s~^3S_1$ and $5s5p~^3P_2\rightarrow5s5d~^1D_2$ transitions.  

It is worth noting that these longest magic wavelengths are located in the same resonance transition region as the 419.88-nm magic wavelength of the $5s^2~^1S_0\rightarrow 5s5p~^3P_0$ transition. This is a crucial requirement for finding a common magic wavelength to achieve triple magic trapping for both the $5s^2~^1S_0\rightarrow 5s5p~^3P_0$ and $5s^2~^1S_0\rightarrow 5s5p~^3P_2$ clock transitions. Additionally, it should be noted that when $\theta_p=0^{\circ}$ there is no magic wavelength for $5s^2~^1S_0\rightarrow 5s5p~^3P_2~\left|M_{i}\right|= 2$ transition in this range. This is due to the cancellation of the contribution of the tensor and scalar terms of the $5s5p~^3P_2 \rightarrow5s6s~^3S_1$ transition to the dynamic polarizability of the $5s5p~^3P_2~\left|M_{i}\right|=2$ states. As a result, the polarizability of $5s5p~^3P_2~\left|M_{i}\right|=2$  changes slowly near the $5s5p~^3P_2 \rightarrow5s6s~^3S_1$ resonance wavelength, as shown in Figure~\ref{fig3}(c). 

\begin{table}
	\caption[]	{Magic wavelengths of the $5s^2~^1S_{0}\rightarrow5s5p~^3P_{2}$ transition in the case of the magnetic field being vertical to the wave vector $\vec{k}$ with the linearly polarized light, that is $\text{cos}^2\theta_p=1$ and $\mathcal{A} = 0$.}\label{Tab6}
	\begin{ruledtabular}
		\begin{tabular}{cccll}			
			Resonances & $\lambda_{\text{res}}$&  \multicolumn{1}{c}{$|M_i|=0$} &\multicolumn{1}{c}{$|M_i|=1$} &\multicolumn{1}{c}{$|M_i|=2$}\\			 
			\midrule
			$5s5p~{^3P}_{2}\rightarrow 5s6s~^3S_1$ & 508.72\\
			& & 433.31(97) & 442.0(1.4) & \\
			$5s5p~^3P_{2}\rightarrow5s5d~^1D_2$ & 365.06\\
			$5s5p~^3P_{2}\rightarrow5s5d~^3D_1$ & 361.55\\
			&&&& 326.01(2)\\
			$5s5p~^3P_{2}\rightarrow5s7s~^3S_1$ & 325.35\\
			& & 303.97(15) & 304.34(25) & 306.18(25)\\
			$5s5p~^3P_{2}\rightarrow 5s6d~^1D_2$ & 300.23\\
		\end{tabular}
	\end{ruledtabular}	
\end{table}

\begin{figure}[tbh]	
 \centering{
 \includegraphics[width=9cm, height=7 cm]{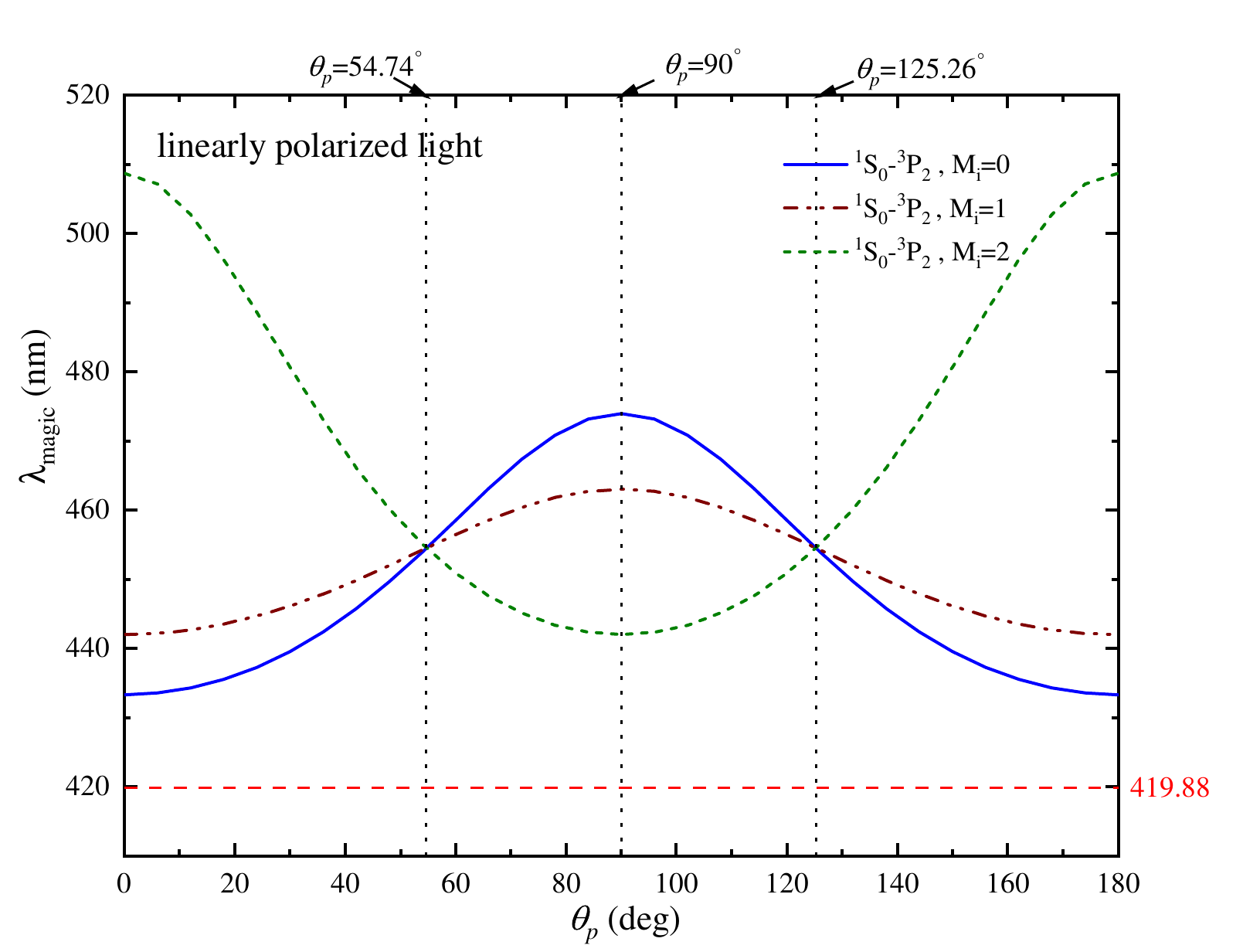} }
 \caption{\label{fig4} The variation of the longest magic wavelengths with the $\theta_p$ under the linearly polarized light for the $5s^2~^1S_0\rightarrow5s5p~^3P_2$ transition.}
 \end{figure}

Figure~\ref{fig4} shows the variation of the longest magic wavelength of the $5s^2~^1S_0\rightarrow5s5p~^3P_2~M_i=0,1,2$ transitions with the angle $\theta_p$.  In the range of $0^\circ \leq \theta_p \leq 90^\circ$, the magic wavelength for the $M_{i}= 0, 1$ transition exhibits a monotonically increasing trend. The minimum values are 433.31 and 442.0 nm at $\theta_p = 0^\circ$ for the $M_i =$ 0 and 1 transitions, respectively. However, for the $M_{i}= 2$ transition, the magic wavelength decreases with increasing angle $\theta_p$, and reaches a minimum value of 442 nm at $\theta_p =90^\circ$. Therefore, all of these magic wavelengths are greater than 419.88 nm. Consequently, it is not possible to find a common magic wavelength that can achieve triple magic trapping for the $5s^2~^1S_0\rightarrow5s5p~^3P_{0}$ and $5s^2~^1S_0\rightarrow5s5p~^3P_2$ transitions.

 \subsection{Magic wavelengths for circularly polarized light}
   
According to Eq.~(\ref{eq4}), the dynamic polarizabilities for the negative $M_{i}$ states with left-handed polarized light are the same as those of the positive $M_{i}$ state with right-handed polarized light. For this reason, in the following discussion we will only give the polarizabilities of the $5s5p~^3P_2~M_i=0,1,2$ states for right-handed polarized light.

 For the right-handed circularly polarized light, $\psi=\pi/4$, $\mathcal{A}=1$, and  $\text{cos}^2\theta_p=\frac{1}{2}\text{sin}^2\theta_{k}$, the Eq.~(\ref{eq4}) can be expressed as:
 \begin{eqnarray}\label{eq11}
 	&&\alpha_i(\omega)=\alpha^S_i(\omega)+ \frac{\text{cos}\theta_k M_{i}}{2J_i}\alpha^V_i(\omega)
 	\nonumber \\
 	&&+\frac{\frac{3}{2}\text{sin}^2\theta_k-1}{2}\frac{3M^2_{i}-J_i(J_i+1)}{J_i(2J_i-1)}\alpha^T_i(\omega),
 \end{eqnarray}
where the polarizability depends on $M_{i}$ and $\theta_k$ due to the contribution of the vector and tensor components, $\alpha^V_i(\omega)$ and $\alpha^T_i(\omega)$.

\begin{figure}[tbh]	
\centering{
\includegraphics[width=9cm, height=7cm]{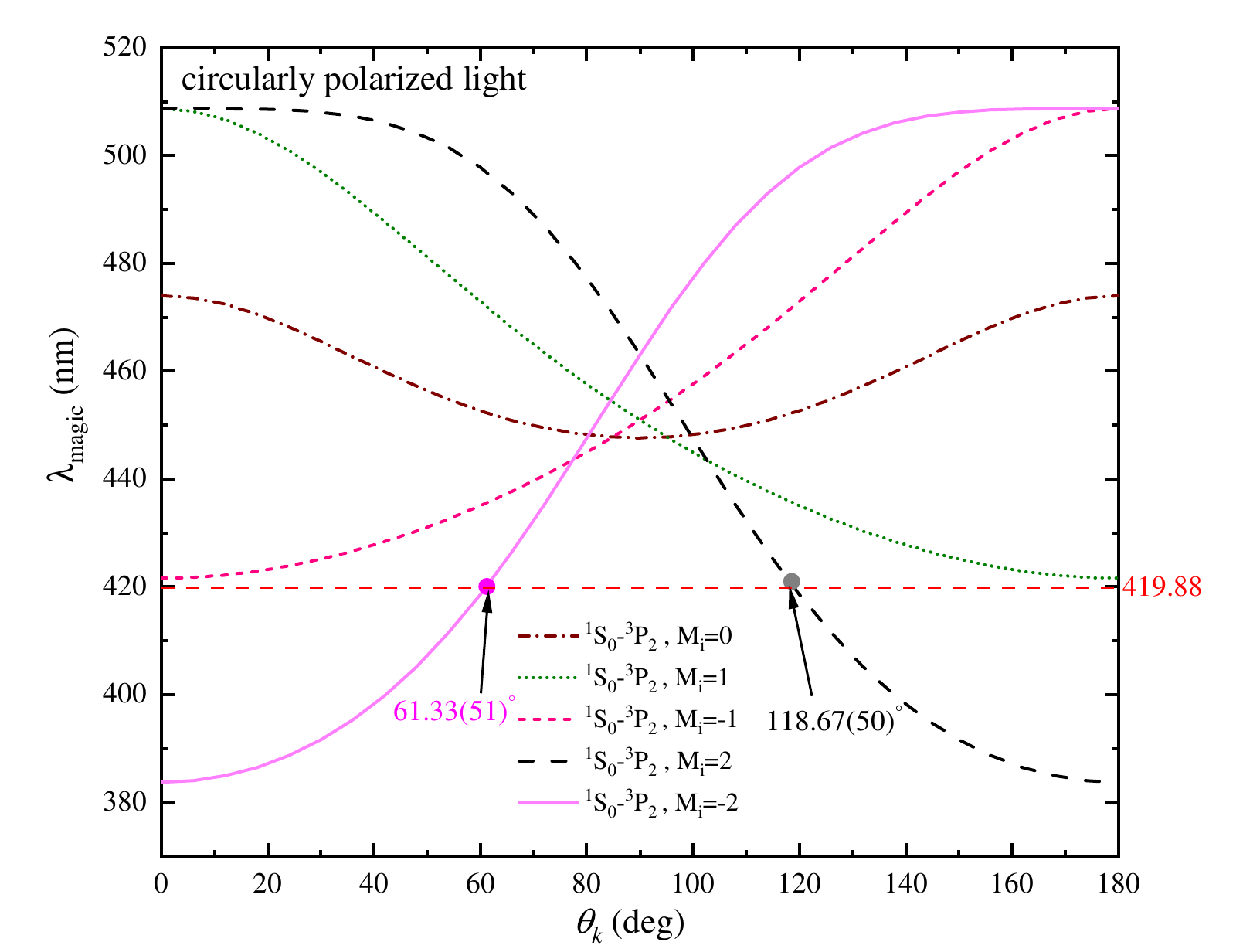} }
\caption{\label{fig5} $\theta_k$-dependent magic wavelengths under the circularly polarized light for the  $5s^2~^1S_0\rightarrow5s5p~^3P_2$ transition. The triply magic trapping conditions are indicated by dots.}
\end{figure}   

When $\theta_k = 90^\circ$, the polarizabilities for circularly polarized light are the same as those for linearly polarized light. Therefore, the magic wavelength should be the same for both cases. Figure~\ref{fig5} depicts the dependence of the longest magic wavelength for the magnetic sublevel transitions of $5s^2~^1S_0\rightarrow5s5p~^3P_2$ $M=0,1,2$ on the angle $\theta_k$.  We can see that the magic wavelengths for the transitions from the $M_i = 0$ and $\pm 1$ states are all greater than 419.88 nm. However, it is worth noting that the magic wavelengths of the $M_i = +2$ and $-2$ transitions are 419.88 nm at $\theta_k=61.33(51)^\circ$ and $118.67(50)^\circ$, respectively. This implies that 419.88 nm is a common magic wavelength for the $5s^2~^1S_0\rightarrow5s5p~^3P_{0}$ and $5s^2~^1S_0\rightarrow5s5p~^3P_2~M_i = \pm 2$ transitions to achieve triple magic trapping.

 \subsection{The triple magic trapping conditions for elliptically polarized light }

\begin{figure}[tbh]	
\centering{
\includegraphics[width=8.5cm, height=8.0cm]{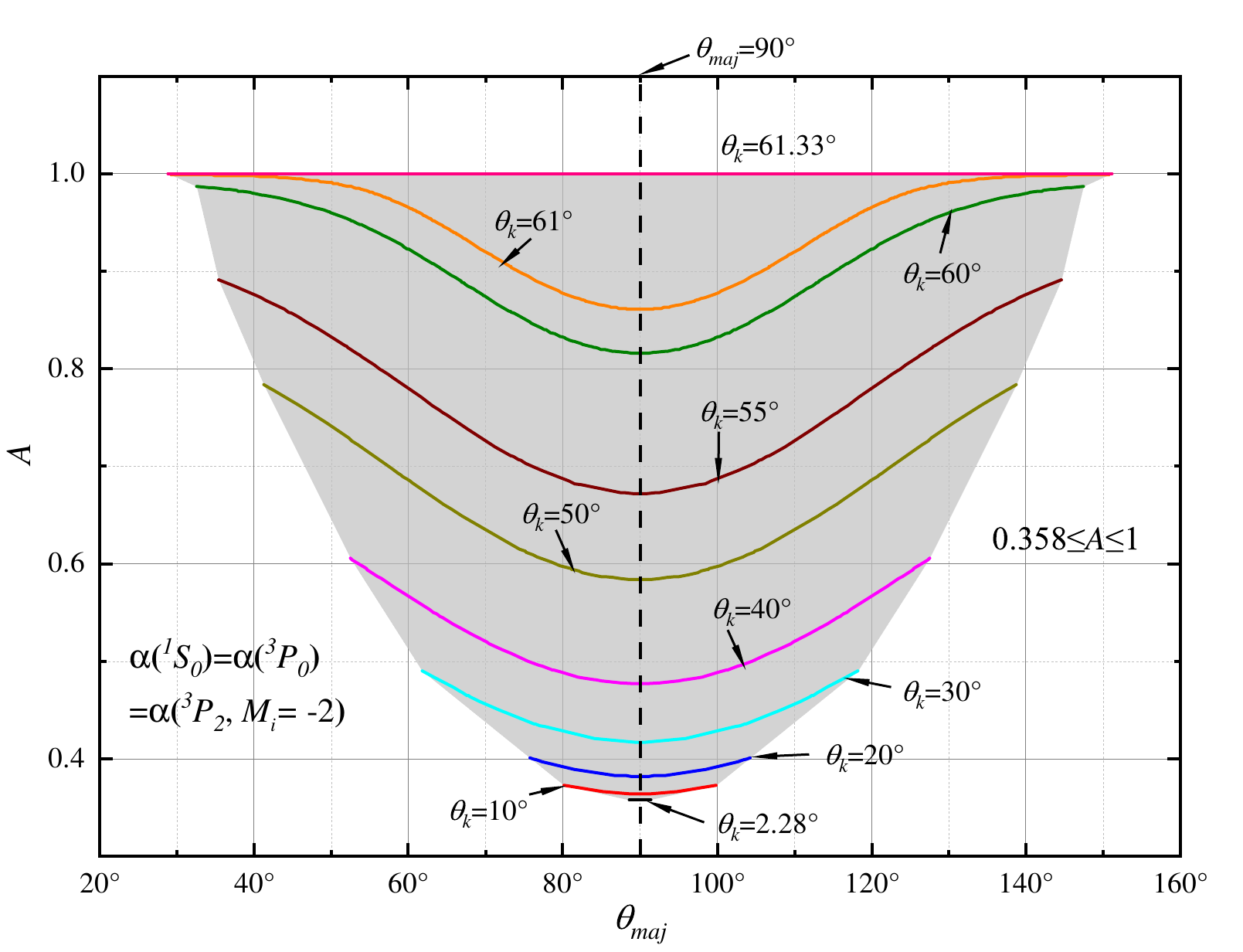} }
\caption{\label{fig6} Triple magic trapping conditions for $5s^2~^1S_0\rightarrow 5s5p~^3P_{0}$ and $5s^2~^1S_0\rightarrow 5s5p~^3P_2~M_i=-2$ clock transitions in the  case of the elliptically polarized light at the 419.88-nm magic wavelength. The gray area represents the three parameters $\mathcal{A}$, $\theta_k$ and $\theta_\text{maj}$ satisfy the conditions of the triple magic trapping.} 
\end{figure}

\begin{figure}[tbh]	
	\centering{
		\includegraphics[width=8cm, height=6.0cm]{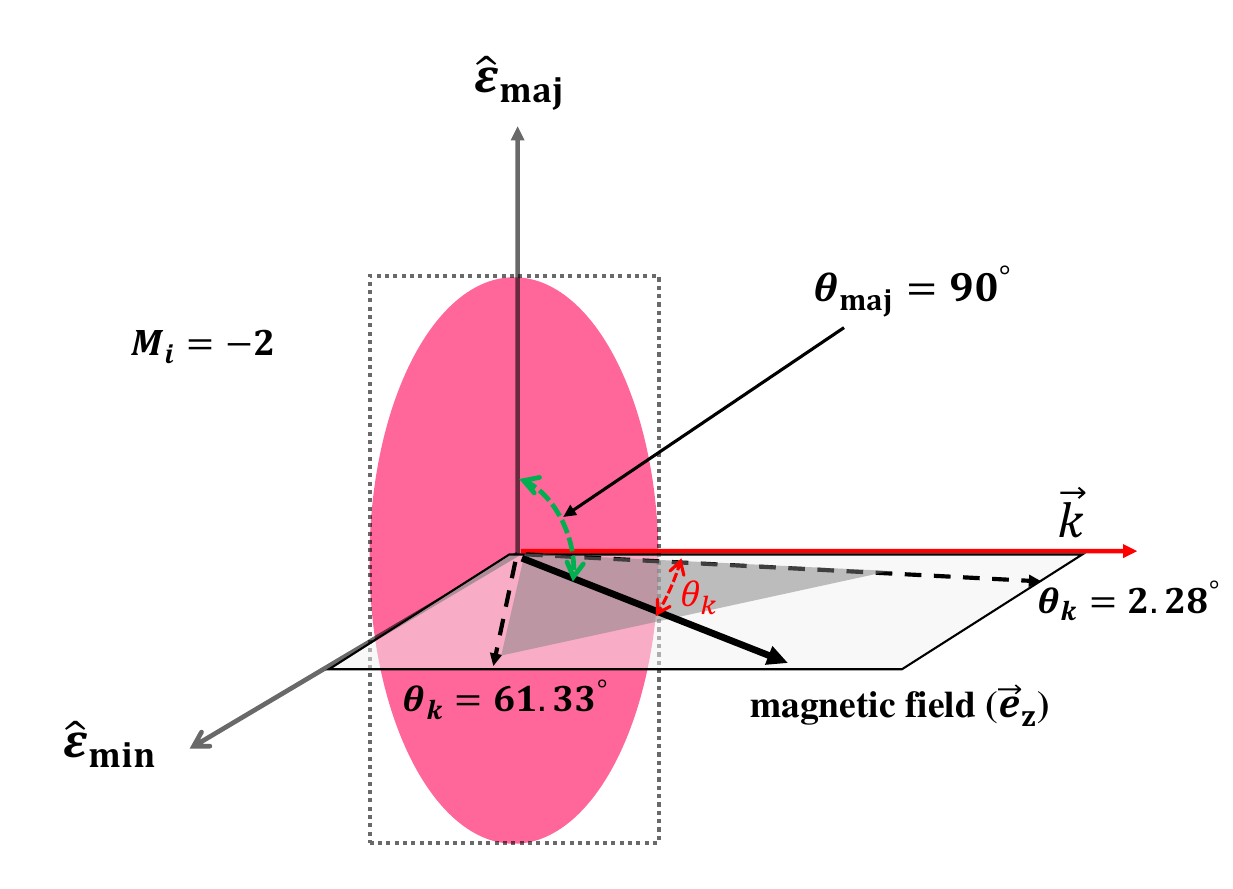} }
	\caption{\label{fig7} Representation of the triple magic trapping conditions when $\theta_\text{maj}=90^\circ$. The magnetic field $\vec{e}_\text{z}$ rotates in the plane composed of $\hat{\varepsilon}_\text{min}$ and $\vec{k}$. The gray area indicates the range of magnetic field directions within which the triple magic trapping condition can be achieved under the elliptically polarized light for the $5s^2~^1S_0\rightarrow 5s5p~^3P_{0}$ and $5s^2~^1S_0\rightarrow 5s5p~^3P_2~M_{i}=-2$ transitions at the longest magic wavelength. }
\end{figure}

According to the Eqs. (\ref{eq4}) and (\ref{eq5}), for the elliptically polarized light, the polarizability of the $5s5p~^3P_2$ state can be expressed as 
    \begin{eqnarray}\label{eq12}
   	\alpha_i(\omega)=&&\alpha^S_i(\omega)+ \mathcal{A}\frac{\text{cos}\theta_k M_{i}}{2J_i}\alpha^V_i(\omega)
   	\nonumber \\
   	&&+\frac{3(\frac{1-\sqrt{1-\mathcal{A}^2}}{2}\text{sin}^2\theta_{k}+\sqrt{1-\mathcal{A}^2}\text{cos}^2\theta_\text{maj})-1}{2}
   	\nonumber \\
   	&&\times\frac{3M^2_{i}-J_i(J_i+1)}{J_i(2J_i-1)}\alpha^T_i(\omega),
   \end{eqnarray}
where, the polarizability for each of the magnetic sublevels depends on $\mathcal{A}$, $\theta _ k$ and $\theta_\text{maj}$. Therefore, the magic wavelength of the clock transition of $5s^2~^1S_0\rightarrow5s5p~^3P_2$ depends on these three parameters as well. 

We have made a thorough analysis of the variation of the longest magic wavelength for each magnetic sublevel transition of the $5s^2~^1S_0\rightarrow5s5p~^3P_2~M = 0,1,2$ with respect to these three parameters. For the $5s^2~^1S_0\rightarrow5s5p~^3P_2~\left|M_i\right|=0,1$ transitions, the magic wavelength is always greater than 420 nm. However, for the $5s^2~^1S_0\rightarrow5s5p~^3P_2~M_i=\pm 2$ sublevel transitions, the variation of the longest magic wavelength is more significant as these parameters change. Interestingly, we find that 419.88 nm is a common magic wavelength for the $5s^2~^1S_0\rightarrow5s5p~^3P_{0}$ and $5s^2~^1S_0\rightarrow5s5p~^3P_2~M_i = \pm 2$ transitions when these three parameters satisfy the relations:
\begin{eqnarray}\label{eq13}
	&&70.9079=76.5011M_i \mathcal{A} \text{cos}\theta_k+32.3004
	\nonumber\\
	&&\times\frac{3(\frac{1-\sqrt{1-\mathcal{A} ^2}}{2}\text{sin}^2\theta_{k}+\sqrt{1-\mathcal{A} ^2}\text{cos}^2\theta_\text{maj})-1}{2},
\end{eqnarray}
where $M_i = \pm 2$. In this equation, the theoretical dynamic polarizabilities $\alpha^S_{^1S_0}(\omega)$, $\alpha^S_{^3P_2}(\omega)$, $\alpha^V_{^3P_2}(\omega)$, and $\alpha^T_{^3P_2}(\omega)$ at $\lambda = 419.88$ nm are used.

In Figure~\ref{fig6}, the gray area represents the range where triple magic trapping at 419.88 nm can be achieved for the $5s^2~^1S_0\rightarrow 5s5p~^3P_0$ and  $5s^2~^1S_0\rightarrow 5s5p~^3P_2~M_i=-2$ clock transitions under right-handed elliptically polarized light. The solid lines indicate the corresponding values of $\theta_k$ for the triple magic trap. It can be seen that triple magic trapping can be achieved within $0.358\leq\mathcal{A}\leq1$, and  $2.28^\circ\leq\theta_k\leq 61.33^\circ$.  For left-handed elliptically polarized light ($\mathcal{A} < 0$), the triple magic condition for the $M_i =-2$ transition is symmetrical to that of right-handed light  ($\mathcal{A} > 0$) with respect to the direction of $\theta_k = 90^\circ$, i.e. $\theta_k$ should satisfy $180^\circ-2.28^\circ\geq\theta_k\geq 180^\circ-61.33^\circ$. For the $M_i =2$ transition, the triple magic condition is the same as for the $5s^2 ~ ^1S_0\rightarrow 5s5p ~ ^3P_2~M_i =-2$, except for the opposite polarization $\mathcal{A}$. 

To make the experiment easier to perform, we suggest setting the angle $\theta_\text{maj}$ to $90^\circ$, i.e. the direction of the magnetic field lies within the plane formed by the $\hat{\varepsilon}_\text{min}$ and $\vec{k}$ axes. In this case, triple magic trapping can be achieved by adjusting only the angle $\theta_k$, as shown in Figure~\ref{fig7}. $\mathcal{A}$ and $\theta_k$ satisfy the following equation:
\begin{eqnarray}\label{eq15}
	87.0581&=&76.5011 M_i \mathcal{A} \text{cos}\theta_k \nonumber \\
       &+&24.2253(1-\sqrt{1-\mathcal{A}^2})\text{sin}^2\theta_k.
\end{eqnarray}
Here, $\mathcal{A}$ should be within the range of $0.358\leq |\mathcal{A}| \leq1$.  For right-handed light, the range for the angle $\theta_k$ is $2.28^\circ\leq\theta_k\leq 61.33^\circ$ (indicated by the gray area in the figure) for the $5s^2 ~ ^1S_0\rightarrow 5s5p ~ ^3P_2~M_i =-2$ transition, and $180^\circ - 2.28^\circ\geq\theta_k\geq 180^\circ- 61.33^\circ$  for the $5s^2 ~ ^1S_0\rightarrow 5s5p ~ ^3P_2~M_i =2$ transition. The range of $\theta_k$ for left-handed light can be determined by its symmetry with right-handed light. 

Finally, we evaluated the sensitivity coefficient~\cite{Flambaum2019,Kozlov2018} for the $5s^2~^1S_0\rightarrow5s5p~^3P_0$ and  $5s^2~^1S_0\rightarrow5s5p~^3P_2$ transitions:
 $q\approx \frac{\omega(+\delta)-\omega(-\delta)}{2\delta}$, where $\omega$ is the transition energy, $\delta$ is the change of the fine structure constant.  
In these estimates, we changed the fine structure constant by 0.1\%, and recalculated the excitation energies of the $5s^2~^1S_0\rightarrow5s5p~^3P_0$ and  $5s^2~^1S_0\rightarrow5s5p~^3P_2$ transitions. The sensitivity coefficients for these two transitions are 5000 and 8150 cm$^{-1}$, respectively. And the enhancement factors~\cite{Flambaum2019,Kozlov2018}, $K=\frac{2q}{\omega}$, are 0.388 and 0.299, respectively. The difference between these sensitivities is small.
  
\section{Conclusions}\label{Sec.IV}

The energy levels and the reduced $E1$  transition matrix elements of Cd atoms are calculated using the RCI+MBPT method. The static and dynamic dipole polarizabilities of the $5s^2~^1S_0$, $5s5p~^1P_1$ and $5s5p~^3P_{0,1,2}$ states are then determined  by the sum-over-states method. The present results are in good agreement with the available theoretical and experimental results. Since the $5s5p ^3P_2$ state is also a long-lived metastable state, the $5s^2~^1S_0\rightarrow5s5p~^3P_2$ transition can also be considered as a second clock transition. The static differential polarizability between the $5s^2~^1S_0$ and $5s5p~^3P_2~M_i=\pm 2$ states is close to that between the $5s^2~^1S_0$ and $5s5p~^3P_0$ states. The magic wavelength for the $5s^2~^1S_0 \rightarrow 5s5p~^3P_0$ and $5s^2~^1S_0 \rightarrow 5s5p~^3P_2$ transitions are identified. The presently calculated magic wavelength of 420.20(57) nm for the $5s^2~^1S_0 \rightarrow 5s5p~^3P_0$ transition is in good agreement with the experimental value of 419.88 nm~\cite{yamaguchi2019} and other theoretical results~\cite{Ye2008}.

In addition, we analyze the common magic wavelengths for the clock transitions $5s^2~^1S_0\rightarrow5s5p~^3P_0$ and  $5s^2~^1S_0\rightarrow5s5p~^3P_2$ under linearly, circularly, and elliptically polarized light. While no common magic wavelength is found for linearly polarized light, it is observed that for  $0.358\leq |\mathcal{A}| \leq1$, 419.88 nm can be used as a common magic wavelength for the $5s^2~^1S_0\rightarrow5s5p~^3P_0$ and $5s^2~^1S_0\rightarrow5s5p~^3P_2~M_i = \pm 2$ transitions to achieve triple magic trapping. We suggest that by setting the angle $\theta_\text{maj}$ to $90^\circ$, triple magic trapping can be achieved by adjusting only the angle between the direction of the static magnetic field and the direction of the wave vector $\vec{k}$ (i.e., $\theta_k$). Overall, our results provide valuable insights into the properties of Cd atoms and the potential for clock transitions at specific wavelengths under different types of light polarization.

\section{Acknowledgments}
This work has been supported by the National Key Research and Development Program of China under Grant No. 2022YFA1602500, the National Natural Science Foundation of China under Grants No. 12174316 and No. 12174268, and the Innovative Fundamental Research Group Project of Gansu Province ( Grants No. 20JR5RA541).

%

\end{document}